\documentclass[twocolumn,times]{aastex62}

\usepackage{graphicx}
\usepackage{epstopdf}
\usepackage{amssymb}
\usepackage[T1]{fontenc}
\usepackage{gensymb}
\usepackage{rotating}
\usepackage{float}
\newcommand{\erg}{${\rm erg \ s^{-1}}$ }
\def\uJy    {$\mu$Jy}
\def\ltsima{$\; \buildrel < \over \sim \;$}
\def\simlt{\lower.5ex\hbox{\ltsima}}
\def\gtsima{$\; \buildrel > \over \sim \;$}
\def\simgt{\lower.5ex\hbox{\gtsima}}

\newcommand{\sfr}{{\rm\,M$_\odot$\,yr$^{-1}$}}

\begin{document}
\begin{sloppypar}
\email{xwshu@mail.ahnu.edu.cn}
\title{Compact Radio Emission from Nearby Galaxies with Mid-infrared Nuclear Outbursts}
\author{B. B.~Dai}
\affiliation{Department of Physics, Anhui Normal University, Wuhu, Anhui, 241002, China} 

\author{X. W.~Shu}
\affiliation{Department of Physics, Anhui Normal University, Wuhu, Anhui, 241002, China}

\author{N. Jiang}
\affiliation{CAS Key Laboratory for Researches in Galaxies and Cosmology, Department of Astronomy, University of Science and Technology of China, Hefei, Anhui 230026, China}

\author{L. M.~Dou}
 \affiliation{Center for Astrophysics, Guangzhou University, Guangzhou 510006, China}

\author{D. Z.~Liu}
\affiliation{Max Planck Institute for Astronomy,  K\"onigstuhl 17,  D-69117 Heidelberg, Germany}

\author{C. W.~Yang}
\affiliation{Polar Research Institute of China, 451 Jinqiao Road, Shanghai, 200136, China}

\author{F. B.~Zhang}
\affiliation{Department of Physics, Anhui Normal University, Wuhu, Anhui, 241002, China}

\author{T. G.~Wang}
\affiliation{CAS Key Laboratory for Researches in Galaxies and Cosmology, Department of Astronomy, University of Science and Technology of China, Hefei, Anhui 230026, China}


\begin{abstract}
We present 5.5 GHz observations with the VLA of a sample of nearby galaxies with energetic 
nuclear outbursts at mid-infrared (MIR) bands. 
These observations reach a uniform depth down to a median rms of $\sim$10\uJy, representing one of most sensitive 
searches for radio emission associated with nuclear transients.  
We detect radio emission in 12 out of 16 galaxies at a level of $>$5$\sigma$, corresponding to a 
detection rate of 75\%. Such a high detection is remarkably different from previous similar searches in 
stellar tidal disruption events.  
{ The radio emission is compact and not resolved for the majority
  of sources on scales of $\simlt0.\arcsec5$ ($<$0.9 kpc at $z$$<$0.1).} 
We find the possibility of the star-formation contributing to 
the radio emission is low, but an AGN origin remains a plausible scenario, 
especially for {sources} that show evidence of AGN activity in their optical spectra. 
If the detections could 
represent radio emission associated with nuclear transient phenomenon such as jet or outflow, 
we use the blast wave model by analogy with the GRB afterglows to describe the evolution of radio light curves. 
In this context, the observations are consistent with a decelerating jet with an energy of $\sim10^{51-52}$ erg 
viewed at 30\degree-60\degree off-axis at later times, {suggesting that powerful jets 
may be ubiquitous among MIR-burst galaxies}. 
Future continuous monitoring observations will be crucial to decipher the origin of 
radio emission through detections of potential flux and spectral evolution. 
Our results highlight the importance of radio observations to constrain the nature of nuclear MIR outbursts in galaxies. 

\end{abstract}


\section{INTRODUCTION}\label{sec:intro}

{It is generally accepted that super massive black holes (SMBH) reside at the cores of many galaxies 
\citep[see review by][]{Graham2015}. When a star passes too close to a SMBH,} it can be squeezed and torn apart once the tidal force of the SMBH exceeds the star's self-gravity \citep{Stone2019}. It is the so-called tidal disruption event (TDE), during which roughly half of the mass of the disrupted star falls back close to the event horizon of black hole, generating a luminous flare of electromagnetic radiation  
\citep{Rees1988,Guillochon2014}. 
As the stellar debris gets accreted effectively, a fraction of accretion power could be 
converted into outflow, leading to the formation of a relativistic jet, which can be detected  
at radio wavelengths \citep[e.g.,][]{vanVelzen2011a,Zauderer2011}. 

\begin{deluxetable*}{cccccccccccc}[!htbp]
\tablecaption{Radio observation results for the sample of galaxies with MIR bursts\label{tab:candidate}}
\tablecolumns{9}
\tablenum{1}
\tablewidth{0pt}
\tablehead{
\colhead{Name} &  \colhead{RA} & \colhead{DEC} &
\colhead{Rms} & \colhead{Integrated flux} & \colhead{Peak flux} & 
\colhead{Beam Size (PA)\tablenotemark{a}} &
\colhead{Host Type} & \colhead{Redshift (z)} & log$L_{\rm [OIII]}$ & $F_{\rm 3GHz}$ & $\Delta$t\\
\colhead{} &  &   & \colhead{$\mu$Jy} & \colhead{$\mu$Jy} & \colhead{$\mu$Jy/beam} & 
\colhead{arcsec $\times$ arcsec(deg)} & \colhead{} & \colhead{} & \colhead{\erg} &$\mu$Jy &{years} 
}
\startdata
SDSSJ1537+5814  & { 234.2970} & { 58.2389 }& 5.8 & 673.1  & 667.7  & 0.32 $\times$ 0.25 (-14.28) & AGN & 0.093 & 40.78 &{ 712} & { 2.15} \\
SDSSJ1340+1842  & 205.1353 & 18.7051 & 11 & $<$33   &  $\dots$  & $\dots$ & AGN & 0.090  &  41.09 & $<$414 & 3.79 \\
SDSSJ1422+0609\tablenotemark{b} & 215.7254 & 6.16483  & 11 & 205  & 205.7  & 0.30 $\times$ 0.28 (-3.86) & Galaxy & 0.038 & $\dots$ &$<$417 & 2.70\\
SDSSJ1133+6701  & 173.4830 & 67.0186 & 25 & 4544 & 4569 & 0.48 $\times$ 0.25 (86.77) & AGN & 0.039 & 39.43 & 4050 &3.83 \\
SDSSJ1549+3327  & 237.4799 & 33.4644 & 13 & $<$39  &  $\dots$  & $\dots$ & Galaxy/SF & 0.086 & 39.68 & $<$330 &2.74 \\
SDSSJ1357+4423  & 209.3621 & 44.3874 & 3.4 & 54.2   &  62.7  & 0.29 $\times$ 0.25 (-4.87) & Galaxy/SF & 0.072 & 40.78 &$<$546 & 4.26\\
SDSSJ1328+6227  & 202.0482 & 62.4619 & 4.4 & 186.2  &  186 & 0.52 $\times$ 0.25 (-85.32) & AGN & 0.092 & 41.34 &$<$405 &3.30 \\
SDSSJ1550+5330  & 237.6914 & 53.5092 & 12 & 1278 & 1279 & 0.52 $\times$ 0.25 (81.91) & Galaxy & 0.066 & 40.87 &748 &4.54 \\
SDSSJ1632+4416  & 248.1951 & 44.2718 & 13 & $<$39 & $\dots$ & $\dots$ & Galaxy & 0.058 & 39.23 & $<$354&2.24 \\
SDSSJ1524+5314  & 231.1588 & 53.2496 & 10 & 204  & 192.3  & 0.31 $\times$ 0.25 (10.39) & Galaxy/SF & 0.085 & 39.78 & $<$318& 3.29 \\
SDSSJ1657+2345  & 254.3617 & 23.7578 & 13 & 97   & 80.5  & 0.36 $\times$ 0.28 (86.51) & AGN & 0.059 & 40.47 &$<$366 &2.66 \\
SDSSJ2217-0820  & 334.3995 & -8.3450 & 5.1 & 135.3  & 131.9  & 0.38 $\times$ 0.27 (-20.88) & AGN & 0.085 & 40.92 & $<$480&3.35\\
SDSSJ1217+0350  & 184.4094 & 3.84455 & 13 & 91  &  88.5 & 0.51 $\times$ 0.28 (-59.83) & Galaxy & 0.073 & 40.33 &$<$411 &4.59 \\
SDSSJ1314+5116  & 198.7493 & 51.2725 & 8.6 & 91.2   & 88.1  & 0.31 $\times$ 0.25 (-34.12) & AGN & 0.025 & 39.93 & $<$339&3.93\\
SDSSJ1156+3131\tablenotemark{c}  & 179.1602 & 31.5200 & 32 & $<$96 & $\dots$ & $\dots$ & Galaxy/SF & 0.080 & 40.94 & $<$873&2.42\\
SDSSJ1402+3922  & 210.5885 & 39.3700 & 4.5 & 117  & 107.1  & 0.38 $\times$ 0.26 (-84.11) & Galaxy/SF & 0.064 & 40.70 &$<$360 &4.35\\
\enddata 
\caption{$^{\rm a}$beam size from the cleaned image. $^{\rm b}$The source is in the SDSS spectroscopy catalog, but the spectrum is not available possibly due to fiber problems. We adopted its photometric redshift instead. $^{\rm c}$The relatively large rms could be due to the contamination from a nearby brighter source. 
}
\end{deluxetable*}

Over the past years, dozens of TDE candidates are discovered at soft X-rays \citep{Komossa2015,Saxton2019}, 
ultraviolet wavelengths \citep[UV;][]{Gezari2009,Gezari2012} and the number is expected to 
increase rapidly 
in the epoch of time-domain optical surveys \citep{vanVelzen2011b, Cenko2012, Arcavi2014, vanVelzen2020}. 
In contrast, radio follow-up observations of known TDEs have resulted in 
very few conclusive detections of jet emission \citep{Bower2013, vanVelzen2013}. 
So far, {very few} jetted TDEs have been discovered with associated radio emission \citep{Zauderer2011, 
Brown2017}, including the prototypical source, Swift J1644+57, in which the radio emission 
is explained due to the interaction of the jet with the ambient medium \citep{Giannios2011, Berger2012, Eftekhari2018}. 
IGR J12580+0134 and Arp 299-B AT1 are found with a relativistic jet that is viewed off-axis \citep{Irwin2015,Perlman2017, 
Mattila2018}, and for the latter a revolved, expanding radio jet is directly imaged. 
{The radio emission has also been detected from the thermal TDE ASASSN-14li \citep{vanVelzen2016a, Alexander2016} 
and XMMSL1 J0750-85 \citep{Alexander2017}, but with luminosities at least two orders of magnitude lower than Swift J1644+57, 
which can be explained by less energetic jets or non-relativistic outflows. 
It is suggested that the outflows maybe more ubiquitous than jets in TDEs \citep{Alexander2016, Anderson2019}, 
but most eluded detections due to insufficient sensitivity of radio observations. }
%
The current non-detections could also be explained by  
the delayed onset of the radio emission at early times. 
Yet the known TDEs with radio detections  
are still not enough to test these possibilities.


In a gas-rich circumnuclear environment, nuclear flares from TDEs can ionize and heat gas 
surrounding the black hole. When dust is present, the UV/optical photons are absorbed by dust 
and re-emit in the infrared (IR). This dust emission is predicted to peak at 3-10 $\mu$m with an 
IR luminosity of $10^{42-43}$ \erg and last for a few years for a typical TDE \citep{Lu2016}. 
Indeed, using archival data at 3.4 and 4.6 $\mu$m (W1 and W2 hereafter) from the WISE all-sky survey, 
such a dust-reprocessed emission is detected in a few TDE candidates 
\citep{vanVelzen2016b, Jiang2016, Dou2016, Dou2017}
providing a new opportunity in diagnosing the physical conditions 
of the circumnuclear material. 
Furthermore, the mid-infrared (MIR) dust echoes from nuclear transients are promising to reveal the population of 
dust-enshrouded TDEs \citep{Mattila2018}, allowing for a more unbiased census of TDE phenomenon. 
{In addition to TDEs, modern time-domain spectral observations have revealed a class of ``changing-look" 
active galactic nuclei (CL AGNs), which can change their optical types on time scales of years, characterized by 
emerged or disappeared broad emission lines as well as dramatic continuum variations \citep[e.g.,][]{Shappee2014, Gezari2017, 
Yan2019}. More importantly, it has been proved that MIR variability is efficient in identifying candidates of CL AGNs \citep{Sheng2017}. 
The radio properties are crucial to understand the nature of CL AGNs, but are yet largely unexplored in observations. }



In this Letter, we report the results from the radio observations of a sample of galaxies with nuclear MIR outbursts, 
revealing a high detection rate of 75\% down to about 60\uJy~ ($\approx5\sigma$). 
The observations and data reductions are described in Section 2. 
In Section 3, we present the properties of radio emission and discussions. 
In Section 4, we summarize our results and findings. 
We adopt a cosmology of $\Omega_{\rm M}=0.3$, $\Omega_{\lambda}=0.7$, and $H_0=70$ km s$^{-1}$ Mpc$^{-1}$ 
when computing luminosity distance.  


\begin{figure*}[htbp]
\begin{center}
\includegraphics[width=7in]{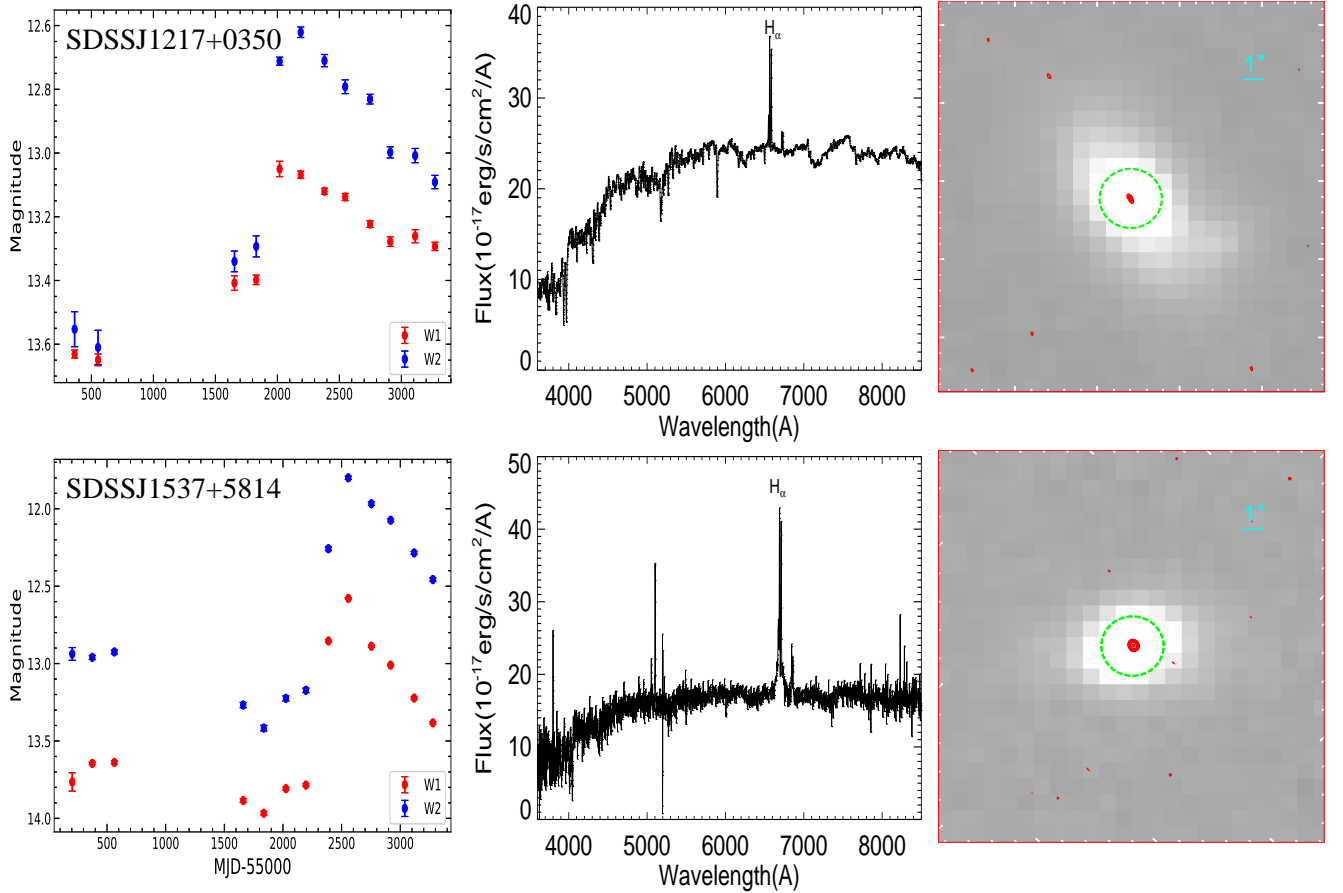}
\end{center}
\caption{Multiwavelength characteristics of two typical sources in our sample, with one optically classified 
as normal galaxy (top) and one as AGN (bottom). From left to right, the WISE MIR light curve at W1 (red) and W2 (blue) band, 
preflare SDSS optical spectrum of host galaxy, and SDSS $r$-band image overlaid with the VLA continuum contours (red). 
The green circle corresponds to the fiber size of SDSS spectroscopy, with a diameter of 3$\arcsec$. The image has a size of $18\arcsec\times18\arcsec$. }
\label{fig:vla_ims}
\end{figure*}

\section{Observation and Data reduction} \label{sec:data}
Our sample of objects are selected from the on-going program to systematically identify 
 MIR outbursts in nearby galaxies (MIRONG, Jiang et al. submitted) 
from the WISE survey. 
We restricted the sample to include only sources not detected in the previous FIRST and NVSS survey, in order 
to reduce the possibility of radio-loud AGNs causing the transient IR emission. 
In addition, we require all objects to be nearby at $z<0.1$. 
The final sample for VLA observations consists of 16 sources, with 9 being optically 
classified as star-forming or normal galaxies and 7 as AGNs based on the SDSS spectroscopy\footnote{One source in the sample, SDSSJ1422+0609, is listed in the SDSS spectroscopy catalog and classified 
as galaxy, but its spectrum is not made public.}. 
Note that we did not exclude the AGNs since recent works have suggested that the 
expected TDE rate in AGNs maybe higher than the quiescent galaxies \citep[e.g.,][]{Karas2007}, and indeed the 
pre-existing AGN activity is found present {in a few TDE candidates 
\citep[e.g.,][]{Blanchard2017, Shu2018, Liu2020}.} 
In addition, the inclusion of AGNs allows us to examine whether the radio detection rate is 
different as compared with normal galaxies.

\begin{figure*}[htbp]
\centering
\begin{minipage}{0.97\textwidth}
\centerline{\includegraphics[width=1.0\textwidth]{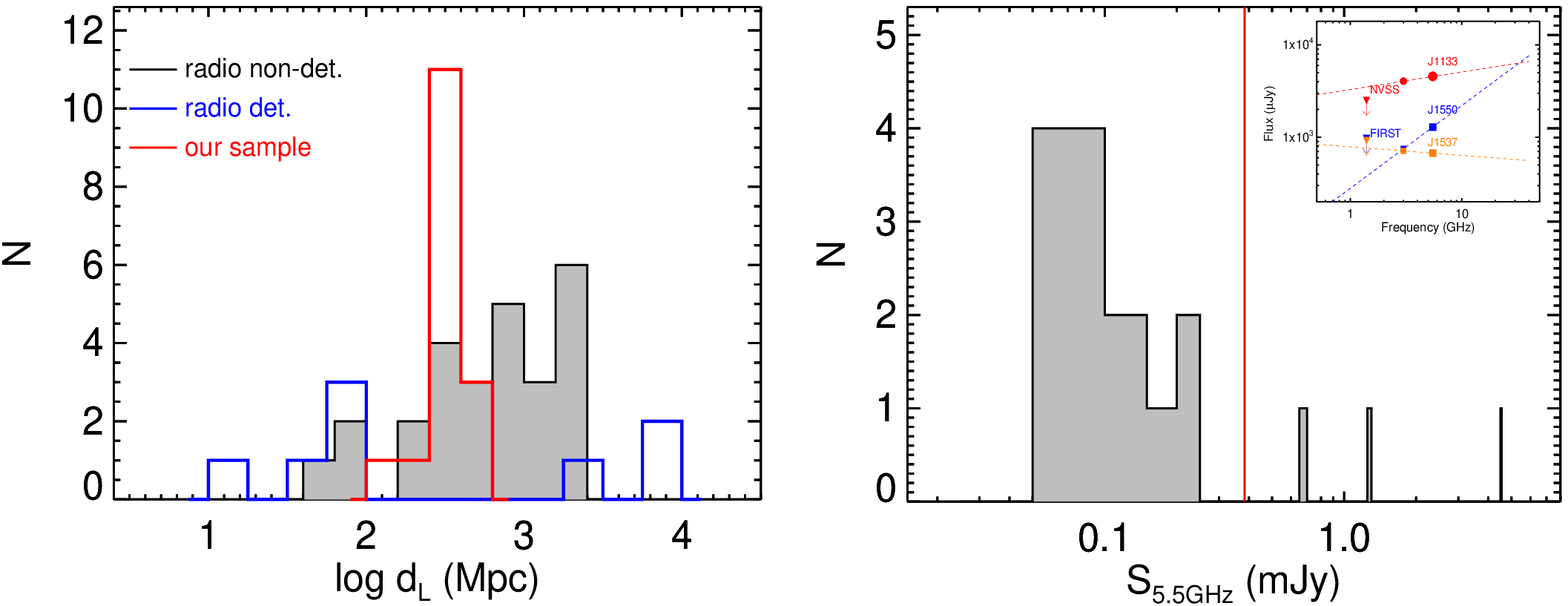}}
\end{minipage}
\caption{
{\it Left panel:} Luminosity distance (redshift) distribution as compared with the TDE sample with and without radio 
detections from literature compilation \citep{Berger2012, vanVelzen2016a, Alexander2017, Blagorodnova2017, Brown2017, Perlman2017, Mattila2018, Anderson2019, Nicholl2019, vanVelzen2019, Wevers2019, Gomez2020, Holoien2020}.
{\it Right panel:} Distribution of flux densities as observed with 
VLA. The red vertical line represents the nominal 5$\sigma$ upper limit of FIRST survey converting 
to 5.5GHz assuming a radio spectral index $\alpha =-0.7$. 
Inset panel shows the radio flux {and spectral slope} for the three brightest sources in the sample, as compared with 
the FIRST/NVSS upper limits. 
}
\label{fig:flux_redshift}
\end{figure*}

All radio observations are performed with Jansky Very Large Array (VLA) in A configuration (program code: 18A-207) 
and at C-band with a total of bandwidth 2 GHz over the frequency range 4.5 $\sim$ 6.5 GHz, consisting of 16 subbands with 
64 channels in each subband. We refer to the central frequency of 5.5 GHz in the following analysis.  
All observations are performed between March 9 and April 30 2018, and each source has a 
total integration time of 25 minutes, with 10 minutes on-source and 15 minutes for calibrators. 
We selected the most suitable radio source close to target for 
amplitude and phase calibration, and 3C286 or 3C48 for bandpass and flux density calibration. 
The data were reduced following standard procedures with the CASA package. 
According to the pipeline log files, we examined each spectral window and 
flagged the abnormal data due to RFI or hardware issues. 
The calibrated data for each source were then selected 
and we used the {\tt CLEAN} algorithm to remove possible contamination from side-lobes,  
with the conventional Briggs weighting and ROBUST parameter of 0. 
The final cleaned maps have a typical synthesized beam of 0.\arcsec3-0.\arcsec5 and a rms noise of $\sim$5-20 \uJy.  

{In addition to the new VLA C-band data, we also checked for the detectability of radio emission 
of the sample from the new VLA Sky Survey (VLASS) at S-band ($\approx$3 GHz), in which the quick-look 
imaging data are available\footnote{https://archive-new.nrao.edu/vlass/quicklook/}. 
The observations were performed between 2017 and 2019, with a median rms of 138 \uJy/beam. 
We found 3 out of 16 galaxies are detected at the S-band. }
In Table \ref{tab:candidate}, we summarize the results of radio observations for the 
16 sources. The columns represent (1) source name; 
{(2) }\& {(3) optical coordinate;} (4) image rms near the source position, $\sigma$; (5) integrated flux density, in $\mu$Jy; 
(6) peak flux density, in $\mu$Jy beam$^{-1}$; (7) image beam size and position angle, 
in arcsecond and degree; (8) the host galaxy type as classified from the SDSS optical spectra
\footnote{http://skyserver.sdss.org/dr16/en/tools/quicklook/summary.aspx?}; (9) redshift $z$, and 
(10) optical {\sc [O iii]} line luminosity in \erg; {(11) integrated flux density at $S$-band 
from the VLASS; (12) The time interval between the start of IR flare 
and radio observations.} 
For non-detections, we give a 3$\sigma$ upper limit to the flux density, which is in the range $\sim$40 to 100 \uJy.  

\section{RESULTS and discussions} \label{sec:results}


\begin{figure*}[htbp]
\centering
\begin{minipage}{0.97\textwidth}
\centerline{\includegraphics[width=1.0\textwidth]{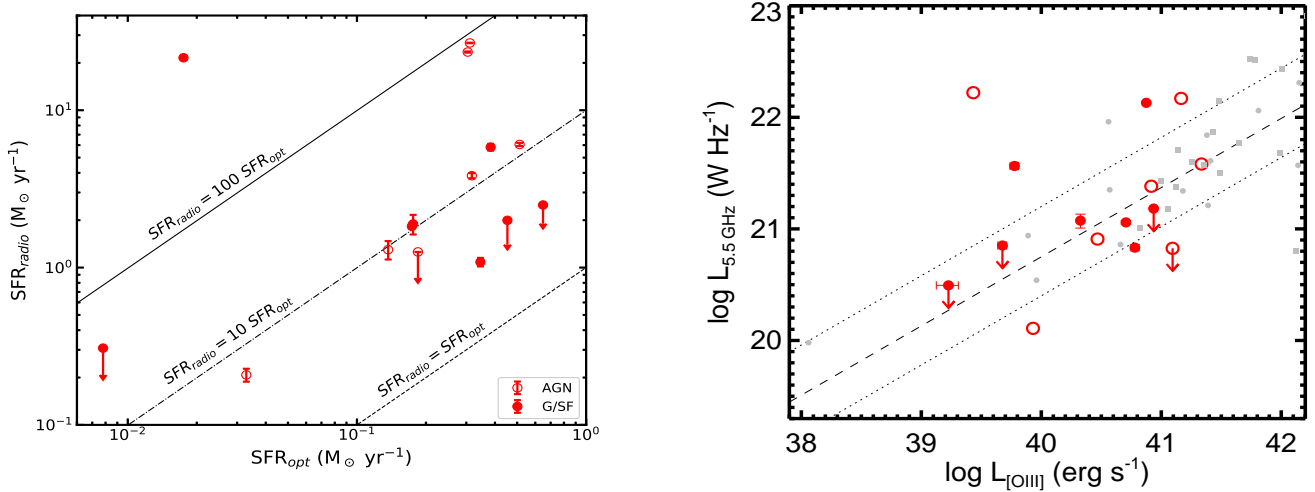}}
\end{minipage}
\caption{
	{\it Left panel:} Comparison of radio-inferred SFRs with the SFRs obtained from SDSS fiber spectroscopy. 
	The galaxy subsample is shown in solid circles, while 
	the AGN subsample is shown in open circles. 
	{\it Right panel:} Radio luminosity versus the extinction-corrected {\sc [O iii]}$\lambda5007$ luminosity. 
	The dashed line represents the fit to radio-quiet Seyfert galaxies \citep{Ho2001}. 
}
\label{fig:sfr-radio-op}
\end{figure*}

As shown in Table \ref{tab:candidate}, we detected radio emission in 12 out of 16 galaxies observed with VLA, with flux 
down to 60$\mu$Jy ($\approx$5$\sigma$), i.e., a 75\% detection rate. The detection rate in the subsample of AGNs ($\approx$86\%) 
is slightly higher than that in galaxies ($\approx$67\%). 
Figure 1 displays the WISE lightcurve, SDSS optical spectroscopy and SDSS $r$-band image overlaid
with radio contours for one star-forming galaxy and one AGN from the sample, respectively.
We argue that the radio emission for the galaxy subsample 
may not be dominated by the AGN activity.  
We constructed a comparison sample from galaxies without MIR variability in the SDSS Stripe 82 
region, where deep 1.4 GHz observations over 92 deg$^2$ were performed with VLA. 
The VLA survey has an angular of $\approx$1.8\arcsec with a median rms noise of 52 \uJy/beam, and 
is the deepest survey achievable to date over an area of $\sim100$ deg$^2$ \citep{Smolvcic2017}. 
Assuming a typical radio spectral index of $\alpha=-0.7$ ($S_{\nu}\propto\nu^{\alpha}$, Smol{\v{c}}i{\'c} et al. 2017), 
the sensitivity of the VLA Stripe 82 survey 
corresponds an rms of $\sim$20 \uJy/beam at 5.5 GHz, comparable to our observations. 
To ensure similar host properties as our sample, the galaxies from Stripe 82 were matched to each source 
in redshift with $\left|\Delta z\right|\le0.005$ and 
WISE 22$\mu$m magnitude with $\left|\Delta mag\right|\le0.05$. 
We selected randomly five matched galaxies for each source, constituting of a comparison sample 
with totally 80 galaxies. 
These galaxies were then cross-matched with the VLA Stripe82 catalog \citep{Hodge2011}, yielding only 
three radio detections, i.e, a detection rate of $<$5\%. 
This indicates that the radio emission in our sample is possibly associated with the nuclear MIR bursts.
{Furthermore, we have compiled SDSS AGNs at z<0.1 from the Stripe 82 survey, 
and matched them to the VLA data within the same area. 
We found radio detections in 61 out of 342 matched sources, i.e., 17.8\% detection rate.  
Hence, the radio detection rate in our IR-burst galaxies seems to be higher 
than that of normal AGNs. }

Most TDE candidates discovered in the optical and soft X-ray bands have no conclusive 
radio detections.  \citet{vanVelzen2013} performed deep search for radio emission in 
seven thermal TDEs, and none of them is detected though the sensitivity of $\sim$10 \uJy~is 
comparable to ours. \citet{Bower2013} searched for the late-time radio emission from seven 
X-ray selected TDE candidates and detected two with a flux level of $\sim$100 \uJy. 
However, the radio identification for one source is ambiguous and the observed radio emission 
for another may be due to a highly variable AGN \citep{vanVelzen2016a}. At a given sensitivity, the previous non-detections may 
result from a distance effect. In Figure 2 (left), we show the luminosity distance (converted 
from redshift) distribution of our sample, and the well-studied TDE candidates in literature 
with and without radio detections for comparison. 
We found that except jetted TDEs where the relativistic beaming effect is important, all 
previous radio-detected objects are within a redshift $z=0.02$ (or $d_{\rm L}\simlt$90 Mpc). 
{Conversely, for the 26 TDE candidates without radio detections, only three at $z<0.02$ and 
most at $z>0.1$ with a median redshift of $z=0.146$ ($d_{\rm L}=692$ Mpc). For comparison, all our sources at 
$z<0.1$, with a median redshift of $z=0.072$, seeming to fill the gap in redshift distribution 
between previous observations with and without radio detections. 
However, it should be noted that there is clear overlap between our sample and the TDEs 
that are not detected at radio.  
While the sensitivity is comparable to ours, we find half of them (those at $z<0.1$) to have radio observations 
performed within 100 days since the optical discovery. 
In this case, the radio non-detections could be simply due to the delayed onset of jet/outflows 
at later times, which can be tested with future monitoring observations. 
Hence, we conclude that the distance effect can play a role, but is not enough to 
explain all radio non-detections. 
 }


The ratio of integrated flux to peak flux is in the range 0.86-1.2, 
with a median value of 1.03, suggesting that most, if not all, radio emission is unresolved and extremely compact 
($\simlt$0.\arcsec5). 
We checked that the positional offsets of radio sources relative to the optical centers are in the range 
0.\arcsec03-0.\arcsec21, with a median offset of 0.\arcsec07, indicating that the radio emission originates from the 
nuclear region. 
The radio flux distribution is shown in Figure 2 (right). The median flux for the 12 detected sources 
is $186\pm4.4$ \uJy, and the source with brightest radio emission is SDSS J1133+6701 (J1133), 
which has an integrated flux density of $S_{\rm 5.5 GHz}=4.5\pm0.03$ mJy. 
{  
In addition to J1133, two other sources in the sample, SDSS J1550+5330 (J1550) and SDSS 1537+5814 (J1537), 
have brighter radio flux, which have also been detected by the VLASS at the S-band. 
Although the VLASS observations are not quasi-simultaneous (about half year earlier) 
as ours, we attempted to measure the radio spectral index between the S-band and C-band for 
the three sources, as shown in Figure 2 (right, inset panel). 
We found that the radio slope is either flat or inverted, with $\alpha$ in the range $-$0.09 to 0.89, 
suggesting the radio emission is from an optically thick region, likely in an early phase of evolution. 
More interestingly, using the spectral index, we extrapolated the VLA 5.5 GHz flux to 1.4 GHz, yielding 
a flux $S_{\rm 1.4 GHz}=3.5$ mJy, which is a factor of 1.4 higher than the 5$\sigma$ upper limit of 
$S_{\rm 1.4 GHz}=2.5$ mJy/beam obtained from the NVSS survey.  
Considering the much larger beam size (by two orders of magnitude) of the NVSS survey,
the expected flux difference is even larger.
The results confirm that at least the radio emission for J1133 is transient (increased in flux). }

Thanks to the high spatial resolution of the VLA observations, 
we have demonstrated that the radio emission is extremely compact as originating from nuclear region. 
We begin to consider whether the detected radio emission could be due to 
circumnuclear star formation process primarily from high-mass X-ray binaries and diffuse hot gas 
heated by supernovae. 
Under the assumption of star formation as origin for the radio emission, 
we calculate the radio-inferred star formation rates (SFRs) based the radio flux,  
using the expression given by \citep{greiner2016}, 
\begin{equation}
{\rm SFR_{radio}} = 0.059\, {\rm M_{\odot} \ yr^{-1}}\, F_{\rm \nu,\mu Jy}d_{L,{\rm Gpc}}^2\nu_{\rm GHz}^{-\alpha}(1+z)^{-(\alpha + 1)}
\end{equation}
where $F_{\nu}$ is the observed flux density at a frequency $\nu$, $d_L$ is the 
luminosity distance at a source redshift. 
Since the expression is extrapolated from the relation between SFR and 1.4 GHz luminosity, 
we adopt a typical value of $\alpha=-0.7$ to account for proper $k-$corrections at 5.5 GHz. 
On the other hand, we cross-matched for each source with the SFRs as measured from the optical emission 
lines (mainly  {\sc [O ii]}) within the 3\arcsec~given by the MPA-JHU SDSS spectroscopic catalog\footnote{https://www.sdss.org/dr12/spectro/galaxy\_mpajhu/}. 
Figure \ref{fig:sfr-radio-op} displays the comparison of radio-inferred SFRs with that obtained from optical emission lines. 
It can been seen that except two objects, the radio-inferred SFRs are about an order of magnitude higher than the SDSS spectroscopic 
SFRs. 
Thus, we consider a star formation origin for the radio emission to be unlikely (with a contribution less than 10\%). 
A single, young supernova seems also impossible to explain the observed radio emission, given the 
low SFRs in the nucleus for most of our sources ($\simlt0.3$\sfr).
In addition, the radio luminosities at 5.5 GHz are higher than the majority of the radio supernova 
remnant ($L_{\nu}\simlt10^{20}$ W Hz$^{-1}$) such as those studied in the starburst galaxy Arp 220 \citep{Varenius2019}. 



Because of its compactness, the radio emission may be associated with a persistent AGN activity 
{or extreme AGN variability due to transient BH accretion}, 
especially for sources that show evidence of an AGN from optical emission lines (50\%). 
None of sources are detected in the ROSAT survey, with an X-ray luminosity $L_{\rm 0.2-2 keV}<10^{41}$\erg, 
suggesting either intrinsic weak or heavily obscured in the soft X-rays. 
It is suggested that optical {\sc [O iii]}$\lambda5007$\AA~luminosity ($L_{\rm [O III]}$) is one of the best measures of 
the intrinsic luminosity of the nuclei of AGNs if present. 
Assuming all the power is due to AGN, we show the relation between $L_{\rm [O III]}$ and radio luminosity 
in Figure 3 (right). 
The dashed line is the best fit to radio-quiet Seyfert galaxies from Ho \& Peng (2001) for which 
the luminosities are plotted in grey dots. We also included the measurements from the radio-quiet 
narrow-line Seyfert 1s \citep[filled grey squares,][]{Berton2016, Berton2018}. 
The dotted line represents the approximately 1$\sigma$ scatter at the luminosity range. 
{While the majority of our sources (including upper limits) are consistent with the Ho \& Peng (2001) relation 
within the 1$\sigma$ scatter, there are four sources with more than 0.5 dex higher flux than the expected. }
However, we note that since very few of the radio-quiet Seyferts from previous works have 
$L_{\rm [O III]}<10^{41}$\erg as the majority of our sources, the relation is very uncertain 
at low $L_{\rm [O III]}$. 
Based on the present data, we cannot rule out the weak AGN origin for the radio emission. 
Actual detections of uniform radio flux variability are required to distinguish the origin from 
a transient accretion phenomenon or a weak persistent AGN. 








\begin{figure}[htbp]
\centering
\begin{minipage}{0.42\textwidth}
\centerline{\includegraphics[width=1\textwidth,angle=270]{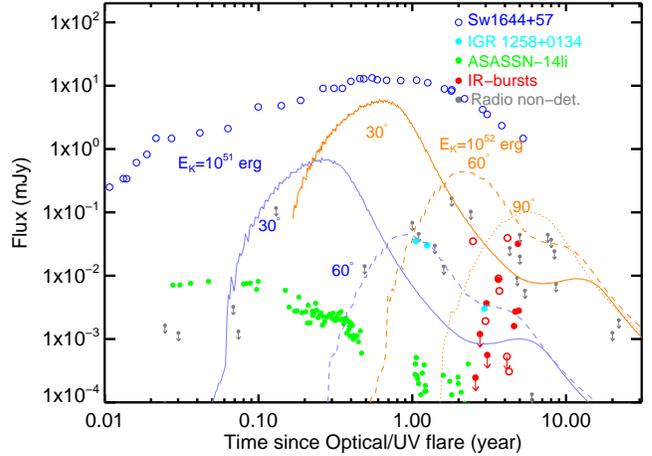}}
\end{minipage}
\setlength{\abovecaptionskip}{-0.5cm}
\caption{
Radio flux at $\approx$5 GHz is plotted with respect to the time of central optical/UV flare. 
For our sample, we assumed a typical distance of dust echo of 0.1pc \citep{vanVelzen2016b} to 
estimate the approximate time of radio observations since central flare. 
The symbol has the same coding as in Figure 3. 
Radio light curve of the jetted TDE Sw 1644+57, the TDE IGR1258 with off-axis jet, ASASSN14li with 
non-relativistic outflow, and upper limits for other optical/X-ray TDEs {(grey dots, 
the same as shown in Figure 2, left) }
are also shown for comparison. 
All flux densities are scaled to the luminosity distance of Sw 1644+57.
The orange and blue lines represent the blast wave model in analog with the 
GRB afterglow as calculated from the {\tt BOXFIT} code (see the text).
}
\label{fig:off-axis}
\end{figure}

We next consider whether the radio emission could be produced by a relativistic jet or outflow 
presumably due to the transient BH accretion, and its interaction with the circumnuclear medium (CNM). 
This is possible since our sample by selection is rich of dust and gas in the circumnuclear region. 
However, any initially relativistic jet (if presents) would have decelerated to non-relativistic velocities 
by the time of our radio observations. 
Similar process has been invoked in gamma-ray burst (GRB) with external shocks from the 
decelerating blast wave potentially contributing to the radio emission. 
We attempted to place constraints on the presence of a relativistic jet by generating a grid of the GRB 
afterglow models with the {\tt BOXFIT} code \citep{vanEerten2012}. 
{Strictly speaking, the blast-wave GRB jet model is not appropriate for AGN jets, 
so any conclusion should be treated with caution. }
In these models, the expansion of jet in a uniform medium is calculated using 
relativistic hydrodynamical simulations.   
We assumed a CNM with constant density medium $n=1$ cm$^{-3}$, 
a jet opening angle of $\theta_{\rm jet}=0.1$, and the fraction of energy in the electrons 
and magnetic fields, $\epsilon_{e}=0.1$ and $\epsilon_{B}=0.01$, respectively. Then we used  
two intrinsic jet energies, $E_{K}=10^{51}, 10^{52}$ erg to generate light curves at different viewing angles. 
Light curves observed at 30\degree, 60\degree and 90\degree from the jet axis are shown in Figure 4. 
For comparison, we also plot the light curve of Sw J1644+57 with an on-axis relativistic jet, 
IGR J12580+0134 with a jet viewed off-axis and ASASSN-14li with a non-relativistic outflow, respectively. 
Since our sample selection is based on the dust reprocessed emission, 
we assumed a typical dust echo distance of 0.1pc \citep{vanVelzen2016b} to infer the time of radio observations 
with respect to the central optical/UV flare, which is around 2-5 years. 
We find that at least for the three sources with brightest radio emission, a jet as powerful as Sw 1644+57 
viewed at 30\degree-60\degree off-axis could yield consistent results with our observations. 
The radio emission for fainter sources could be consistent with the less energetic jet, 
similar to the off-axis jet observed in IGR J12580+0134, though a non-relativistic outflow 
is also possible if formed at a later time. 
Because the radio emission (if transient) is expected to continue declining in the next few years, 
multi-epoch monitoring of these sources at radio bands is highly encouraged for comparison with models in detail.     

{On the other hand, comparing with the upper limits (5$\sigma$) of most optical/X-ray TDEs (grey dots), 
our late time radio observations are likely revealing that either the jet emission as powerful as IGR J12580 
is ubiquitous among IR-selected outburst galaxies, or it is associated with a new weak-AGN variability mode. 
Future more sensitive late time radio observations of optical/X-ray TDEs will be helpful to test these possibilities. 
If the radio emission is due to TDEs, our radio observations of IR-selected flares have demonstrated that it is 
promising to reveal a new population of dust-obscured TDEs that are largely missed from previous optical surveys 
\citep{Mattila2018}. Studies on the multi-wavelength properties of galaxies with nuclear IR-flares will be presented elsewhere. }

\section{Conclusion} \label{sec:conclusion}

We have presented VLA radio observations of galaxies with nuclear MIR outburst that were discovred in the WISE all-sky survey.  
We detect the radio emission at 5.5 GHz in 12 out of 16 galaxies down {to 5$\sigma$ upper limit of $\sim$60\uJy, }providing 
one of most sensitive searches for radio emission in nuclear transients such as TDEs.   
The majority of sources are unsolved at a resolution of $\simlt0.\arcsec5$, indicating the radio emission is uniformly compact. 
The location of the radio emission is consistent with the optical center of the galaxy, with a positional offsets less than 0.\arcsec2, 
 supporting an origin from nucleus. 
 The SF contribution to the 5.5 GHz emission at a similar spatial scale is found to be low ($\simlt$10\%). 
 {The AGN origin for the radio emission remains a possible scenario with the current data.
 By checking for the VLASS data, we find radio detections in three sources and  
 a flat or inverted radio spectrum between 3 GHz and 5.5 GHz for them, suggesting 
 an origin from optically thick region.  
 We find convincing evidence in one source that the radio emission is transient. 
 }
 If explaining the radio emission as a jet interaction with the ambient medium, 
 by analog with the blast wave model in GRB afterglows, we find the observations to be consistent 
 with a decelerated jet viewed at 30\degree-60\degree off-axis, similar to the one observed in the TDE 
 IGR J12580+0134. 
 We have obtained follow-up observing time from the VLA for several sources in the sample, which 
 will provide measurements of potential radio flux and spectral variability, hence be able to shed new 
 insights into the origin of the radio emission in galaxies with nuclear MIR outbursts.

 \acknowledgments
 
 The authors thank VLA operations staff for their assistance in scheduling and performing the observations. 
 We thank Hendrik van Eerten for kind advice on the use of {\tt BOXFIT} code. 
 The National Radio Astronomy Observatory is a facility of the
 National Science Foundation operated under cooperative agreement
 by Associated Universities, Inc.
 This research made data products from the Wide-field Infrared Survey Explorer. 
 The work is supported by Chinese NSF through grant 11822301, 11833007, U1731104.

\bibliography{radio_mirburst.bbl}

\end{sloppypar}
\end{document}